\documentclass[aps,pra,amsmath,amssymb,amsfonts,superscriptaddress,floatfix,nofootinbib]{revtex4} 
\pdfoutput=1
\usepackage{amsmath,amsthm,amsfonts,amssymb}
\usepackage{graphicx}

\usepackage{color}
\usepackage{hyperref}

\newtheorem{conjecture}{Conjecture}

\newtheorem{lemma}{Lemma}

\newtheorem{definition}{Definition}

\usepackage{algorithm}
\usepackage[capitalize,nameinlink]{cleveref}

\newcommand{\be}{\begin{equation}}
\newcommand{\ee}{\end{equation}}

\newcommand{\expec}{\mathbb{E}}
\newcommand{\lov}{{\vartheta}}

\begin{document}

\title{Limitations and Separations in the Quantum Sum-of-squares, and the Quantum Knapsack Problem}
\author{Matthew B.~Hastings}
\begin{abstract}
We answer two questions regarding the sum-of-squares for the SYK model left open in \cite{hastings2022optimizing}, both of which are related to graphs.  First (a ``limitation"), we show that a fragment of the sum-of-squares, in which one considers commutation relations of degree-$4$ Majorana operators but does not impose any other relations on them, does not give the correct order of magnitude bound on the ground state energy.  Second (a ``separation"), we show that the graph invariant $\Psi(G)$ defined in \cite{hastings2022optimizing} may be strictly larger than the independence number $\alpha(G)$.  The invariant $\Psi(G)$ is a bound on the norm of a Hamiltonian whose terms obey commutation relations determined by the graph $G$, and it was shown that $\alpha(G)\leq \Psi(G) \leq \vartheta(G)$, where $\vartheta(\cdot)$ is the Lovasz theta function.
We briefly discuss the case of $q\neq 4$ in the SYK model.
Separately, we define a problem that we call the quantum knapsack problem. 

Note added: after this paper appeared on the arxiv, it was pointed out to me that  Ref.~\cite{xu2023bounding} previously found examples of graphs where $\Psi(G)>\alpha(G)$.  However, the example presented here seems to be new, and gives a counter-example to conjecture 29 of Ref.~\cite{xu2023bounding}.
\end{abstract}
\maketitle

\section{Introduction and Background}
The sum-of-squares method is a general technique for \emph{lower bounding} the ground state energy of a quantum Hamiltonian.  By decomposing some Hamiltonian $H$
as $H=\sum_\alpha Q_\alpha^\dagger Q_\alpha + \lambda$, we prove that the ground state energy of $H$ is at least $\lambda$.  Here $\lambda$ is a scalar and each $Q_\alpha$ is typically chosen to be a polynomial in some set of operators, such as Majorana operators (for fermionic systems) or Pauli operators (for quantum spin systems).  If the degree of the polynomials $Q_\alpha$ is bounded, then optimal such decompositions can be found by semi-definite programming techniques.

In \cite{hastings2022optimizing}, this method was applied to the SYK model 
~\cite{FW70,BF71,SY93,Kit15,rosenhaus2019introduction,GV16,GJV18,FTW19,FTW18,FTW20} , a model in which the Hamiltonian is a polynomial in Majorana operators, with randomly chosen coefficients.  This polynomial is taken to be homogeneous of some given degree $q$.
Since the SYK model is, in some sense, maximally strongly interacting, this is an interesting testbed for the sum-of-squares method, which can allow access to nonperturbative information.
It was shown that a certain ``fragment" of the degree-$6$ sum-of-squares method\footnote{Here, a degree $2k$ sum-of-squares method means that each $Q_\alpha$ is at most degree $k$ in the Majorana operators.  By a ``fragment", we mean that we incorporate some, but not all, of the power of the sum-of-squares, by restricting the possible $Q_\alpha$ or by using only some subset of the relations between the operators.} was capable of giving the correct order of magnitude bound on the ground state energy of the SYK model, with high probability.

Several questions were left open in that paper.  Here we answer two such questions, both of which are related to a \emph{graphical} interpretation of the sum-of-squares.
We additionally briefly consider the SYK model for $q\neq4$ and describe a problem that we call the quantum knapsack problem.

Before getting into these questions, we review the SYK model and this graphical method.  If this material is new, one should read \cite{hastings2022optimizing} which provides essential background on the SYK model and the sum-of-squares, as we give only a brief review here.

\subsection{SYK Model}
In the SYK model, we have some set of $n$ Majorana operators, labeled $\gamma_1,\ldots,\gamma_{n}$, where $n$ is even.
These operators anti-commute with each other, they are Hermitian, and they square to the identity.  There are no other relations between these operators and this set of relations can be realized by operators on a Hilbert space of dimension $2^{n/2}$.

The SYK Hamiltonian is a homogeneous polynomial of degree $q$ in these operators.
For $q$ even, one writes 
\be
\label{SYKeven}
H=i^{q/2} \sum_{i_1 < i_2 < \ldots <,i_q} J_{i_1,\ldots,i_q} \gamma_{i_1} \ldots \gamma_{i_q}.
\ee
where $J_{i_1,\ldots,i_q}$ is a scalar.  Note that given any monomial of degree $q$ in $q$ distinct Majorana modes, we may re-order it (up to sign) so that the indices appear in increasing order, e.g., we may replace $\gamma_2 \gamma_1$ with $-\gamma_1 \gamma_2$.
For odd $q$, one writes
\be
\label{SYKodd}
H=i^{(q-1)/2}\sum_{i_1 < i_2 < \ldots <,i_q} J_{i_1,\ldots,i_q} \gamma_{i_1} \ldots \gamma_{i_q}.
\ee
We can regard $J_{\ldots}$ as a tensor, which we assume is anti-symmetric in its indices\footnote{By anti-symmetrizing it, we can replace the restricted sum $i_1 < i_2 < \ldots <,i_q$ with an unrestricted sum over all $i_1,\ldots,i_q$ and multiply by a factor of $1/q!$.}, or we may regard it as a vector, indexed by a single index $i_1,\ldots,i_q$.  In either case we denote it $\vec J$.

We can write this as $H=\sum_\alpha J_\alpha O_\alpha$ where $\alpha$ indexes choices of $i_1,\ldots,i_q$, and where $O_\alpha$ is equal to the product  $\gamma_{i_1} \ldots \gamma_{i_q}$ multiplied by a power of $i$ so that it is Hermitian.
The different $O_\alpha$ are monomials in Majorana operators.  
We sometimes write such a monomial as $O_{j_1,j_2,\ldots}$ to indicate that it is monomial $\gamma_{j_1} \gamma_{j_2} \ldots$.
We will say that two monomials $O_{j_1,j_2,\ldots}$ and $O_{k_1,k_2,\ldots}$ \emph{agree} on an odd (respectively, even) number of indices, if there are an odd (respectively, even) number of $j_a$ which also appear as an index $k_b$, i.e., if $\{j_1,j_2\ldots\} \cap \{k_1,k_2\ldots\}$ has odd or even cardinality.

In the SYK model, one chooses the $J_\alpha$ randomly.  They may be chosen to be independent, identically distributed Gaussian random variables with zero mean, such that
the expectation value $\expec[\sum_\alpha J_\alpha^2]$ is equal to $1$.  Alternatively, one may choose $J_\alpha$ to be a uniformly random vector on the sphere such that  $\sum_\alpha J_\alpha^2$ equals $1$ exactly, rather than just in expectation.  In the large $n$ limit, by standard concentration of measure arguments, if
$\expec[\sum_\alpha J_\alpha^2]=1$, then with high probability $\sum_\alpha J_\alpha^2$ is close to $1$.

This normalization that we choose for $\expec[\sum_\alpha J_\alpha^2]$ is \emph{different} from that usually used in the high energy physics literature,
where instead one usually takes the expectation value proportional to $n$.  The reader should be careful of this when comparing to other papers.

With our normalization, with high probability the ground state energy for $q=4$ is $\Theta(\sqrt{n})$.  See \cite{hastings2022optimizing} for proofs of
this and for previous results.

As a point of terminology, we will use ``an instance of the SYK model" to refer to a random choice of the $J_\alpha$, while the ``SYK Hamiltonian"
will refer to either \cref{SYKeven} or \cref{SYKodd} for some (possibly non-random) choice of the $\vec J$.

\subsection{Sum-of-Squares}
Given a Hamiltonian $H$, a sum-of-squares proof of a lower bound on the ground state energy is a decomposition $H=\sum_\alpha Q_\alpha^\dagger Q_\alpha + \lambda$.
By standard semi-definite programming duality, there is a dual to such sum-of-squares proof.  Such a dual consists of a \emph{pseudo-expectation}.   We briefly review this; see
\cite{hastings2022optimizing} for more.
We emphasize for computer scientists that we will be focused on lower bounds for the smallest eigenvalue of an operator, rather than upper bounds for the largest eigenvalue.  So, there will be some sign differences compared to \cite{hastings2022optimizing}

To define such a pseudo-expectation, choose some set of operators $\{O_\alpha\}$.  This set will include all the $O_\alpha$ used to define $H$ by $H=\sum_\alpha J_\alpha O_\alpha$, but may also include additional operators. 
 In particularly, this set will typically be chosen to include the identity operator, $1$. Then, the pseudo-expectation of some product $O_\alpha^\dagger O_\beta$ is written as $\tilde \expec[O_\alpha^\dagger O_\beta]$, where the tilde indicates that this is a pseudo-expectation value.
 For the degree-$2k$ sum-of-squares, this set will be the set of all monomials in Majorana operators of degree at most $k$.
 
 It is convenient to multiply $O_\alpha$ by a phase, if needed, so that $O_\alpha$ is Hermitian.  We do this from now on.

This pseudo-expectation can be written as a matrix, regarding $\alpha$ as a row index and $\beta$ as a column index.  As a matrix, it must be Hermitian and positive semi-semi-definite.  It must also obey some linear constraints.  We will use three different kinds of linear constraints.  First, we use the constraint that certain operators square to the identity, and that the pseudo-expectation value of the identity is $1$.  Thus $\tilde \expec[1^\dagger 1]=1$, and $\tilde \expec[O_\alpha^\dagger O_\alpha]=1$ if $O_\alpha^2=1$.   Indeed, for the operators we use for the SYK model, $O_\alpha^2=1$.   The second kind of linear constraint is that if $O_\alpha$ and $O_\beta$ commute then
$\tilde \expec[O_\alpha O_\beta]=\tilde \expec[O_\beta O_\alpha]$, while if they anti-commute then
$\tilde \expec[O_\alpha O_\beta]=-\tilde \expec[O_\beta O_\alpha]$.
The third kind of linear constraint we use refers to any other relations among Majorana operators, such as a constraint that if $O_\alpha=\gamma_1 \gamma_2 \gamma_3 \gamma_4$
and $O_\beta=\gamma_5 \gamma_6 \gamma_7 \gamma_8$
and $O_\gamma=\gamma_1 \gamma_2 \gamma_3 \gamma_5$
and $O_\delta=\gamma_4 \gamma_6 \gamma_7 \gamma_8$,
then $O_\alpha O_\beta O_\gamma O_\delta=-1$.
In this case we would enforce that $\tilde \expec[O_\alpha O_\beta]=-\tilde \expec[O_\gamma O_\delta]$.
We will refer to the first two kinds of constraints as the commutation and anti-commutation relations, while we will refer to the third kind of constraint as other relations.

Then, the semi-definite programming dual of the sum-of-squares method is to minimize the pseudo-expectation of the Hamiltonian (which is $\sum_\alpha J_\alpha \tilde \expec[1 O_\alpha]$) subject to all these constraints.

\subsection{Graphs, Commutation Relations, and Graph Invariants}
It is convenient to encode these commutation relations in some graph $G$.
This graph has vertex set $V_G$ and edge set $E_G$, with vertices labeled by the same discrete index $\alpha$ that labels the operators $O_\alpha$.  There is an edge between $\alpha,\beta$ if they anticommute and no edge if they commute.  (Of course, this assume that every pair either commutes or anti-commutes but that is the case for the models we consider.)  We assume that $O_\alpha^2=1$ whenever we consider a graphical model.

Then, it was shown \cite{hastings2022optimizing}, that if $\sum_\alpha J_\alpha^2=1$, that a sum-of-squares method of degree $2$ in the $O_\alpha$ (i.e., this is a fragment of the degree-$2q$ sum-of-squares method in the Majorana operators), which uses only the commutation and anti-commutation relations and no other relations between the $O_\alpha$, will have $\tilde \expec[H^2] \leq \vartheta(G)$, where $\vartheta$ is the Lovasz theta function, for \emph{any} choice of $J_\alpha$ with $\sum_\alpha J_\alpha^2=1$.  Indeed, the ``worst case" choice of $J_\alpha$ will be such that $\tilde \expec[H^2] = \vartheta(G)$, where the worst case is defined below

\subsection{Outline}
In the next two sections, we answer two questions left open in \cite{hastings2022optimizing}.
In \cref{crelonly}, we consider the power of the sum-of-squares for the SYK model, and others model, imposing only commutation and anti-commutation relations, and no other relations.  We show that if these are the only relations imposed, then at \emph{any} degree in the sum-of-squares one cannot obtain the correct scaling of the ground state energy with $n$ in the SYK model.  In \cref{graphinvar}, we consider a certain graph invariant, $\Psi(G)$, defined to be the maximum, over $J_\alpha$ with $\sum_\alpha J_\alpha^2=1$, of the square of the operator norm Hamiltonian obeying the commutation and anti-commutation relations of that graph.  It was shown previously that $\alpha(G)\leq \Psi(G)\leq \lov(G)$.  However, no example was known where $\alpha(G)<\Psi(G)$.  Here we construct such an example.  We also make some remarks on the possibility of ``rounding" a solution of the semi-definite program to an actual quantum state.
Then, in \cref{SYKqn4} we briefly discuss the SYK model for $q\neq 4$.
Finally, in \cref{qknapsack} we discuss a problem that we call the ``quantum knapsack problem".

\section{Sum-of-Squares with Only Commutation and Anti-commutation Relations}
\label{crelonly}

In this section, we consider the sum-of-squares method applied to models with only commutation and anticommutation relations.
In \cref{d2graph}, we give some general results on a degree-$2$ sum-of-squares, with commutation relations defined by some graph.  Using these general results, we show that the degree-$8$ (in terms of Majorana operators, so degree-$2$ in terms of degree-$4$ monomials of Majorana operators) sum-of-squares, applied to the $q=4$ SYK model using only commutation and anti-commutation relations, does not give the correct scaling of the ground state energy.
In 
\cref{commrelSYK} we show that at \emph{any} degree of the sum-of-squares, if one uses only commutation and anti-commutation relations, it does not give the correct scaling of the ground state energy for the SYK model.

These two subsections have a different flavor.  \cref{d2graph} uses properties of the semi-definite program.  \cref{commrelSYK} constructs a set of operators which obey the given commutation and anti-commutation relations but no other relations, and then considers the exact ground state energy in this model, which then implies bounds on sum-of-squares using only these relations.

\subsection{Degree-$2$ Sum-of-Squares In Terms of a Graph}
\label{d2graph}

Consider a Hamiltonian $H=\sum_\alpha J_\alpha O_\alpha$, with commutation and anti-commutation relations defined by some graph, and $O_\alpha$ being Hermitian and $O_\alpha^2=1$.

Let us begin by defining two different semi-definite programs.
First, we can bound the expectation value of $H^2$ using the following semi-definite program.  Define a matrix Hermitian $M$ whose rows and columns are also labeled by this same discrete index $\alpha$.  Let the matrix element $M_{\alpha,\beta}$ be the pseudo-expectation value of $O_\alpha O_\beta$.  Then, we impose that $M_{\alpha,\beta}=- M_{\beta,\alpha}$ if $(\alpha,\beta)\in E_G$ and $M_{\alpha,\beta}=+M_{\beta,\alpha}$ otherwise and we impose that $M_{\alpha,\alpha}=1$.

We maximize $\sum_{\alpha,\beta} M_{\alpha,\beta} J_\alpha J_\beta.$  This gives a semi-definite program whose value upper bounds the expectation value of $H^2$.
In fact, given any $M$ which obeys these constraints, the complex conjugate of $M$ (i.e., complex conjugate each entry) also obeys these constraints and has the same value for the objective function.  Hence, given any $M$, we may replace $M$ with its real part to obtain a matrix obeying these constraints with the same value for the objective function.
So, we impose that $M$ is a \emph{symmetric} matrix, with $M_{\alpha,\beta}=0$ if $(\alpha,\beta)\in E_G$.

\begin{definition}
We call the semi-definite program of the above paragraph the ``graph semi-definite program for $H^2$.
\end{definition}

\begin{definition}
We also define another semi-definite program, that we call the ``graph semi-definite program for $H$."
To see this, consider the set of pseudo-expectation values of operators of form $1, O_\alpha,$ or $O_\alpha O_\beta$.  We can write these pseudo-expectation values in a matrix $N$, with one additional row and column compared to $M$.  We have the following block decomposition
$$
N=\begin{pmatrix}
1 & v \\ v & M\end{pmatrix},$$
where $v$ is a real vector of pseudo-expectation values $O_\beta$ and $M$ obeys the same linear constraints as for 
the graph semi-definite program for $H^2$.
The expectation values of $H$ is equal to
$(\vec J,v)$ where $\vec J$ is the vector of coefficients $J_\alpha$.  
We minimize this expectation value, subject to the requirement that $N$ be positive semi-definite.
\end{definition}

Remark:
imposing that $N$ is positive semi-definite imposes also that
$(\vec J,v)^2 \leq \sum_{\alpha,\beta} M_{\alpha,\beta} J_\alpha J_\beta.$
Here we use parentheses $(\cdot,\cdot)$ to denote an inner product.
As a result, the square of the optimum of the graph semi-definite program for $H$ is less than or equal to the
optimum of the graph semi-definite program for $H^2$.

\begin{definition}
The worst case of the graph semi-definite program for $H^2$ is defined to be the maximum of the optimum of this semi-definite program over all $\vec J$ subject to $|\vec J|^2=1$.

Similarly, the worst case of the graph semi-definite program for $H$ is defined to be the minimum of the optimum of this semi-definite program over all $\vec J$ subject to $|\vec J|^2=1$.
\end{definition}

Following\cite{hastings2022optimizing},
consider the worst case of the graph semi-definite program for $H^2$.
For given $\vec J$ and given $M$, we may definite a new matrix $M'$ by
$M'_{\alpha,\beta}=M_{\alpha,\beta} J_\alpha J_\beta$.  So, for given $\vec J$, the graph semi-definite program
for $H^2$ is equivalent to a semi-definite program with variable $M'$ where we maximize the sum of entries of $M'$, subject to
$M'_{\alpha,\beta}=0$ if $(\alpha,\beta)\in G_E$ and $M'_{\alpha,\alpha}=J_\alpha^2$.
For any matrix $M'$, there is a $\vec J$ with $|\vec J|^2=1$ such that $M'_{\alpha,\alpha}=J_\alpha^2$ iff $\sum_\alpha M_{\alpha,\alpha}=1$.
So, the maximum over all choices of $\vec J$ is given by solving the following semi-definite program:
maximize the sum of entries of $M'$, subject to
$M'_{\alpha,\beta}=0$ if $(\alpha,\beta)\in G_E$ and subject to $\sum_\alpha M_{\alpha,\alpha}=1$.
The value of this semi-definite program is called the Lovasz theta-function of $G$, and we write it $\lov(G)$.

\begin{lemma}
\label{lemVtrans1}
Suppose $G$ admits some vertex-transitive group action (recall that this means that for any pair of vertices, some element of the symmetry group maps the first vertex to the second).
Then, the worst case of the graph semi-definite program for $H$ is equal to minus the square-root of the worst case of the graph semi-definite program for $H^2$.

Further, this worst case for both semi-definite programs can be achieved with a $\vec J$ that is proportional the all $1$s vector.
\begin{proof}
Consider the semi-definite program for $M'$ above, 
where we maximize the sum of entries of $M'$, subject to
$M'_{\alpha,\beta}=0$ if $(\alpha,\beta)\in G_E$ and $M'_{\alpha,\alpha}=J_\alpha^2$.
We may conjugate any feasible $M'$ by a permutation corresponding to any element of the symmetry group of $G$, and the result is still a feasible $M'$ with the same value of the objective.
By taking an $M'$ which optimizes the semi-definite program and averaging over such permutations, we get an $M'$ which optimizes the semi-definite program which is invariant under the symmetry group,
so that all diagonal entries are the same.
Since all diagonal entries are the same, this $M'$ corresponds to 
a matrix $M$ which is a feasible solution of the graph semi-definite program for $H^2$, with
$M'_{\alpha,\beta}=M_{\alpha,\beta} J_\alpha J_\beta$ and $\vec J$ proportional to the all $1$s vector.

We now consider the graph semi-definite program for $H$, taking the given $\vec J$ which is proportional to the all $1$s vector.
Consider a feasible solution $N$ of the form
$$
N=\begin{pmatrix}
1 & v \\ v & M\end{pmatrix},$$
where $M$ is as in the above paragraph and is invariant under the symmetry group.
We take $v$ proportional to the all $1$s vector, and minimize $(\vec J,v)$.
Previously, we used that the fact that $N$ is positive semi-definite implies that
$(\vec J,v)^2 \leq \sum_{\alpha,\beta} M_{\alpha,\beta} J_\alpha J_\beta$.  Now we use that $v$ is an eigenvector of $M$ so that in fact we may take equality
$(\vec J,v) = -\sqrt{\sum_{\alpha,\beta} M_{\alpha,\beta} J_\alpha J_\beta}=-\sqrt{\lov(G)}$.
\end{proof}
\end{lemma}

We claim the following:
\begin{lemma}
\label{lemVtrans2}
Suppose $G$ admits some vertex-transitive group action.
Then, the optimum of the graph semi-definite program for $H$ for any given $\vec J$ is less than or equal to
$$-\Bigl(\sum_\alpha |J_\alpha|\Bigr) |V_G|^{-1/2} \sqrt{\lov(G)}.$$
\begin{proof}
Following the proof of \cref{lemVtrans1}, there is a feasible solution $N_0$ of the graph semi-definite program for $H$ such that $v$ is proportional to the all $1$s vector.  In this feasible solution, each entry of $v$ is equal to
$-|V_G|^{-1/2} \sqrt{\lov(G)}$.

Let $D$ be a diagonal matrix of the same size as $N$.  The diagonal entries of $D$ are $(1,\vec\sigma)$, where $\vec\sigma$ is a vector
of the sign of elements of $J_0$: $\vec \sigma_\alpha$ is equal to $\pm 1$ depending on whether $(J_0)_\alpha\geq 0$ or $(J_0)_\alpha<0$.
Then, $DND$ is also a feasible solution of the semi-definite program, and at that point the objective function is
equal to $-\Bigl(\sum_\alpha |J_\alpha|\Bigr) |V_G|^{-1/2} \sqrt{\lov(G)}.$
\end{proof}
\end{lemma}

As an application of \cref{lemVtrans2}, consider the quartic SYK model with $n$ Majorana modes.  In this case, the appropriate graph has a vertex transitive symmetry group.  The symmetry group here is permuting the labels of the Majorana operators: take any permutation $\pi$ on $\{1,\ldots,n\}$ and permute $\gamma_a \rightarrow \gamma_{\pi(a)}$, which then induces some permutation on monomials in the Majorana operators, leaving the commutation relations unchanged.
With high probability, $\sum_\alpha |J_\alpha|=\Theta(n^2)=\Theta(|V_G|^{1/2})$, where $\Theta(\cdot)$ is big-O notation.
So, by \cref{lemVtrans2}, with high probability
the graph semi-definite program for $H$ has an optimum that is less than or equal to
$-\Theta(\sqrt{\lov(G)})$.
At the same time, the worst case of the graph semi-definite program for $H$ is $-\sqrt{\lov(G)}$, so with high probability the optimum indeed \emph{is} $-\Theta(\sqrt{\lov(G)})$.

As shown previously\cite{hastings2022optimizing}, the Lovasz theta function for this graph is asymptotically proportional to $n^2$ for large $n$ and so $H=-\Theta(n)$ in worst case, and further with high probability the
graph semi-definite program for $H$ has an optimum that is $-\Theta(n)$.
However, for typical $\vec J$, one may use degree-$6$ sum-of-squares\cite{hastings2022optimizing} to show that
$H\geq -O(\sqrt{n})$.
 Thus, the graph semi-definite program for $H$ is polynomially weaker than the degree-$6$ sum-of-squares, even though the graph semi-definite program includes terms up to degree $8$, as this graph semi-definite program does not have any of the other relations among products of Majorana operators, only the commutation and anti-commutation relations.

\subsection{Commutation and Anti-Commutation Relations Are Not Enough for SYK}
\label{commrelSYK}
It may be surprising that the graph semi-definite program for $H$ gives such weak bounds for the optimum.  As we have noted, the weakness of this semi-definite program is that it only knows whether two given monomials in the Majorana operators commute or anti-commute.  One may wonder: suppose we considered higher order semi-definite programs which still only knew the commutation and anti-commutation relations.  Would these give better bounds?

The answer is no.  In fact, we will show that if one only imposes the commutation and anti-commutation relations, and no other relation, then an exact solution will, with high probability, give a ground state energy $-\Theta(n)$.

Precisely, what we mean is the following.  
We begin with the case $q=4$.
Let $O_\alpha$ be a set of ${n \choose 4}$ operators which square to the identity.  Each $\alpha$ is a tuple $j_1,j_2,j_3,j_4$ with $1\leq j_1<j_2<j_3<j_4\leq n$, so we write the operators as $O_{j_1,j_2,j_3,j_4}$.  Impose the relation that $O_{j_1,j_2,j_3,j_4}$ commutes (respectively, anti-commutes) with $O_{k_1,k_2,k_3,k_4}$ if  $|\{j_1,j_2,j_3,j_4\} \cap \{k_1,k_2,k_3,k_4\}|$ is even (respectively, odd), i.e., if they agree on an odd number of indices.
Impose no other relations.
Choose $\vec J$ to have independent random Gaussian coefficients, so that $|\vec J|=1$ in expectation.
Then, we claim that with high probability the ground state energy is $-\Theta(n)$.

Indeed, we may define operators $O_\alpha$ that obey these commutation and anti-commutation relations and no other relations as follows.  Take $n$ Majorana modes and ${n\choose 4}$ qubits, labeling the qubits by the same indices $\alpha$.  Let $Z_\alpha$ be the Pauli $Z$ operator on qubit $\alpha$.  Let
$$O_{j_1,j_2,j_3,j_4}\equiv Z_{j_1,j_2,j_3,j_4} \gamma_{j_1} \gamma_{j_2} \gamma_{j_3} \gamma_{j_4}.$$

We can understand the fact that we impose no other relations as follows.  Consider any product $P$ of operators $O_\alpha$,  Since each $O_\alpha$ commutes or anti-commutes with any other $O_\beta$, we may, up to an overall sign, commute the operators in the product $P$ so that they appear in sequence $P=O_{\alpha_1}^{n_1} O_{\alpha_2}^{n_2} \ldots$ where $\alpha_1<\alpha_2<\ldots$ given some arbitrary ordering of the $\alpha$.  Then, since each $O_\alpha$ squares to the identity, we may assume that each power $n_a$ is equal to $0$ or $1$.  Hence, we may just consider products $P$ in which each $O_\alpha$ appears at most once.  There is a trivial product, in which $P$ simply is the identity operator.  However, any nontrivial product $P$ is some product of Majorana operators multiplied by at least one $Z_\alpha$.
Now, consider some nontrivial product $P$ so which is \emph{central}, so that it commutes with all $O_\alpha$.
In this case, it equals a product of at least one $Z_\alpha$, possibly multiplied by the fermion parity operator (i.e., by the product of all Majorana operators).  Hence, one may say that imposing no other relations means that if we have some product of Majorana operators which is a scalar, we have ``forgotten" what the sign of that scalar is.

Then, for random $\vec J$, the ground state energy of $H$ is given by minimizing over all choices of signs $Z_\alpha$, the ground state energy of the quartic SYK Hamiltonian with coupling constants $J'_\alpha=J_\alpha Z_\alpha$, i.e., we can arbitrarily choose the sign of the coupling constants.

For the rest of this section, we will use the notation that $O_\alpha$ refers to the operators with only commutation and anti-commutation constraints, i.e., $O_{j_1,j_2,j_3,j_4}\equiv Z_{j_1,j_2,j_3,j_4} \gamma_{j_1} \gamma_{j_2} \gamma_{j_3} \gamma_{j_4},$ while we will use $O^{SYK}_\alpha$ to refer to the operators from the SYK Hamiltonian,
$O^{SYK}_{j_1,j_2,j_3,j_4}\equiv \gamma_{j_1} \gamma_{j_2} \gamma_{j_3} \gamma_{j_4}$.

Let's give an example first of a specific choice of $\vec J$, which we write as $\vec J^0$, before showing our claim about the ground state energy.  Recall that the Hadamard matrix $H_m$ is an $m$-by-$m$ matrix with $\pm 1$ entries, and with all eigenvalues equal to $\pm\sqrt{m}$.  
For even $n$, let $Q_n$ be the block matrix
$$Q_n \equiv \frac{i}{\sqrt{n/2}} \begin{pmatrix} 0 & H_{n/2} \\  - H_{n/2} & 0 \end{pmatrix}.$$
Since $Q_n$ is anti-symmetric and pure imaginary we can consider a free fermion Hamiltonian
$$H_{\rm free}\equiv i\sum_{i<j} \gamma_i \gamma_j (Q_n)_{i,j}.$$
In the ground state of $H_{\rm free}$, correlation functions obey Wick's theorem, and
we have
$\langle \gamma_i \gamma_j \rangle= \delta_{i,j}-(Q_n)_{i,j}.$
The ground state energy of $H_{\rm free}$ is $-\Theta(n)$.  
Now, choose $J^0_\alpha$ as follows.  Recall each $\alpha$ is a tuple $j_1,j_2,j_3,j_4$.  Choose $$J^0_{j_1,j_2,j_3,j_4}=-\frac{1}{n} \Bigl( Q_{j_1,j_2} Q_{j_3,j_4}\pm{\rm permutations}\Bigr).$$
Here, by $Q_{j_1,j_2} Q_{j_3,j_4} \pm {\rm permutation}$, we mean adding all $4!$ permutations of the indices $j_1,\ldots,j_4$, with a sign given by the sign of the permutation.
Note that the $\ell_2$ norm of $Q_n$ is equal to $\sqrt{n}$, and so the $\ell_2$ norm of
this $\vec J^0$ is asymptotically constant, and so we can rescale $\vec J^0$ by a constant to make the $\ell_2$ norm asymptotically equal to $1$.

Then,
$$\sum_\alpha J^0_\alpha O^{SYK}_\alpha=-\frac{1}{n} H_{\rm free}^2+{\rm subleading},$$
where ${\rm subleading}$ denotes quadratic terms in the Majorana operators which are subleading in $n$.
Evaluating the expectation value of $\sum_\alpha J^0_\alpha$ in the ground state of $H_{\rm free}$, using Wick's theorem, this expectation value is $-\Theta(n)$.
Hence, the ground state energy of $\sum_\alpha J^0_\alpha$ is $-\Omega(n)$.

Remark: there is nothing special about our choice of the Hadamard matrix here.  We are just picking some matrix with $\pm 1$ entries such that the sum of absolute values of eigenvalues is $\Theta(n^{3/2})$ and so that the corresponding $H_{\rm free}$ has ground state energy $-\Theta(n)$.  One could instead use, for example, a matrix where signs of entries are chosen $\pm 1$ independently and uniformly at random.  However, one could not choose the matrix with all entries equal to $+1$.

Now we show:
\begin{lemma}
In this model where we consider only commutation and anti-commutation relations, for random $\vec J$, with high probability the ground state energy is $-\Theta(n)$.
\begin{proof}
Consider a random $\vec J$.  Multiply each $J_\alpha$ by a sign, if needed, so that $\vec J$ has the same sign structure\footnote{i.e., the same sign structure means that each entry has the same sign.} as $\vec J^0$, giving some $J'_\alpha$.  Evaluate the expectation value of $\sum_\alpha J'_\alpha O^{SYK}_\alpha$ in the ground state of $H^{\rm free}$.
Using Wick's theorem, this expectation value is $-1$ times a weighted sum of absolute values of the entries of $J'_\alpha$.
The expectation value (over random $\vec J$) of this expectation value is $-\Theta(n)$, and by standard concentration of measure arguments, with high probability the expectation value of 
$\sum_\alpha J'_\alpha O^{SYK}_\alpha$ in the ground state of $H^{\rm free}$ is $-\Theta(n)$.
\end{proof}
\end{lemma}

Remark: for any even $q\geq 4$, we may define an analogous model where we only impose commutation and anti-commutation relations of the SYK Hamiltonian and ignore all other relations.  Then, a similar argument would show that with high probability, the ground state energy is $-\Omega(n^{q/4})$.  It was conjectured in Ref.~\cite{hastings2022optimizing} that for arbitrary even $q\geq 4$ the worst case ground state energy of the SYK Hamiltonian is $-\Theta(n^{q/4})$ and further that this is certified by the sum-of-squares using just the commutation and anti-commutation relations (and hence it will hold for the model of this section), and this conjecture was verified for a range of small $q$.  We expect this conjecture is true in general, meaning that for \emph{any} instance of the model in this section the ground state energy is $-O(n^{q/4})$.
So, if this conjecture holds, it follows that
with high probability the ground state energy if we consider just the commutation and anti-commutation relations is $-\Theta(n^{q/2})$.

\section{Graph Invariants}
\label{graphinvar}
Given a graph $G$, let $\alpha(G)$ denote the independence number of $G$, i.e., the  cardinality of a maximum independent set of vertices, where an independent set is one in which no pair of vertices in the set are neighbors.  Let $\vartheta(G)$ denote the Lovasz theta function of $G$.  It is well-known that $\alpha(G)\leq \vartheta(G)$ for all $G$.

In Ref.~\cite{hastings2022optimizing}, a further graph invariant $\Psi(G)$ was defined.  It was proven that
$$\alpha(G)\leq \Psi(G)\leq \vartheta(G),$$
but it was left open whether there were any examples where $\alpha(G)$ is strictly less than $\Psi(G)$.  Here we give such an example, and discuss some theory.

We give only a single example.  Thus, we leave open how large the separation between $\alpha(G)$ and $\Psi(G)$ can be.  Note that it is possible to have $\vartheta(G)$ arbitrarily large for $\alpha(G)=2$.  Indeed, for a graph on $n$ vertices, with $\alpha(G)=2$, the maximum possible value of the Lovasz theta function is $\Theta(n^{1/3})$, as shown in \cite{alon1994explicit}, improving results in \cite{kashin1981systems}.
We conjecture that the same is possible for $\Psi(G)$.

\subsection{Definition of $\Psi(G)$}
Let $O_\alpha$ be Hermitian operators on some Hilbert spacing, obeying
$O_\alpha^2=1$
for all $\alpha$ and obeying commutation and anti-commutation relations defined by some graph $G$.

Then, the invariant $\Psi(G)$ was defined to be the maximum, over all choices of real scalars $J_\alpha$ with $\sum_\alpha J_\alpha^2=1$, of
the square of the operator norm $\Vert J_\alpha O_\alpha \Vert$.
Using semi-definite programming methods, it was shown that $\Psi(G)$ is upper bounded by the Lovasz theta function.

Conversely, it is immediate that $\Psi(G)$ is lower bounded by $\alpha(G)$.   Let $S$ be any maximum independent set.
Let $J_\alpha=1/\sqrt{|S|}$ for $\alpha \in S$ and $J_\alpha=0$ otherwise.  Then, the result follows immediately.

\subsection{Example}
The example we have found with $\Psi(G)>\alpha(G)$ was found by a computer search.  It is an explicit example of a graph on $12$ vertices and edges with $\alpha(G)=2$ but $\Psi(G)\geq 2.01\ldots$  Let $\overline G$ denote the \emph{complement} of $G$.  The adjacency matrix of $\overline G$ is
$$
\begin{pmatrix}
0& 1& 0& 1& 0& 0& 1& 1& 0& 0& 0& 0\\  
1& 0& 1& 0& 0& 0& 0& 0& 1& 0& 1& 0\\  
0& 1& 0& 1& 0& 1& 0& 1& 0& 0& 0& 0\\  
1& 0& 1& 0& 1& 0& 0& 0& 0& 1& 0& 1\\  
0& 0& 0& 1& 0& 1& 0& 1& 0& 0& 1& 0\\  
0& 0& 1& 0& 1& 0& 1& 0& 0& 1& 0& 1\\  
1& 0& 0& 0& 0& 1& 0& 0& 1& 0& 1& 0\\  
1& 0& 1& 0& 1& 0& 0& 0& 1& 0& 0& 1\\  
0& 1& 0& 0& 0& 0& 1& 1& 0& 1& 0& 0\\  
0& 0& 0& 1& 0& 1& 0& 0& 1& 0& 1& 0\\  
0& 1& 0& 0& 1& 0& 1& 0& 0& 1& 0& 1\\  
0& 0& 0& 1& 0& 1& 0& 1& 0& 0& 1& 0
\end{pmatrix}.
$$

Let us explain how we found this example, and why we gave the adjacency matrix of the complement above.  This graph $\overline G$ has $26$ edges.  The requirement that $\alpha(G)=2$ is then simply expressed as requiring that $\overline G$ be triangle free.

We constructed various random triangle-free graphs by a process  that is sometimes called the ``triangle-free process"\cite{bohman2009triangle}.  Here one starts with an empty graph, calling that $G_0$.   At step $i$ of the process, we construct a graph $G_i$ by choosing a uniformly random edge and adding it to graph $G_{i-1}$, subject to the condition that adding this edge does not create a triangle in $G_i$.  If no such edge exists, the process terminates.

Having constructed a triangle-free graph, we then construct a Hamiltonian $H$ given by
$$H=\frac{1}{\sqrt{n}} \sum_\alpha O_\alpha,$$ where we find some representation of the $O_\alpha$ as operators on a finite-dimensional Hilbert space.
We then numerically computed the largest eigenvalue of $H$ and determined whether the square of this eigenvalue was larger than $2$.
We repeated this process until we succeeded.  We did not succeed on any randomly generated graphs of size $11$ but did on a graph of size $12$.  In the example graph, the square of the largest eigenvalue is $2.01\ldots$ and the actual value of $\Psi(G)$ may be larger.

Note that we make no attempt to optimize the $J_\alpha$ in the definition of $\Psi(G)$ but rather simply picked $J_\alpha=1/\sqrt{n}$.  See \cref{cjcoeff} however for some possibility to improve our separation between $\Psi(G)$ and $\alpha(G)$ by considering different coefficients.

We also tried taking $\overline G$ to be either the five vertex cyclic graph $C_5$ (which is triangle-free and has chromatic number 3) or the Gr\"{o}tzsch graph (which is a triangle-free graph on $11$ vertices and has chromatic number 4) but neither one gave an example with $\Psi(G)>\alpha(G)$.  We also tried taking $G$ (instead of $\overline G$) to be $C_5$ and that also did not give such an example.

We make the following conjecture:
\begin{conjecture}
For any real $r>0$, then for all sufficiently large $n$,
if the triangle-free process is run to generate a graph $\overline G$ of size $n$, then 
$$\Vert \frac{1}{\sqrt{n}}\sum_\alpha O_\alpha \Vert \geq r$$ with high probability, and so as a corollary
$\Psi(G)\geq r^2$ with high probability.
\end{conjecture}

Note that it is possible to realize operators $O_\alpha$ which obey these commutation and anti-commutation relations (and no other relations) as acting on a finite dimensional Hilbert space.  In Ref.~\cite{hastings2022optimizing}, a general technique for doing this was given, requiring a Hilbert space of dimension $2^{n-r}$ where $n$ is the number of vertices in $G$. and $2r$ is the $\mathbb{F}_2$ rank of the adjacency matrix of $G$.

Let us give a slightly simpler method of finding a representation.  The method here requires a potentially larger Hilbert space dimension $2^n$.  This is a disadvantage for numerical simulation.  However it has the advantage that the representation can be constructed more simply and so was convenient for a first simulation.  We construct it by taking a system of $n$ qubits.  Each operator $O_\alpha$ is mapped to $X_\alpha \prod_{\beta<\alpha, (\alpha,\beta)\in E} Z_\beta$, where $E$ is the edgeset of $G$, and where we impose some ordering on the vertices of $G$. That is, we map $O_\alpha$ to Pauli $X$ on qubit $\alpha$, multiplied by a product over Pauli $Z_\beta$ over qubits $\beta<\alpha$ such that $O_\alpha$ and $O_\beta$ anti-commute.

We expect that by using instead the representation of Ref.~\cite{hastings2022optimizing} which uses a smaller Hilbert space, and also by using Lanczos or other matrix-vector methods to find the largest eigenvalue of $H$ (rather than doing a full diagonalization as we did) it would be possible to explore significantly larger triangle-free graphs.

\subsection{Weighted Generalization}
\label{cjcoeff}
We remark that there is a natural extension of these conjectures to the case of \emph{weighted} independent set.  Indeed, the weighted extension of the result can be obtained by considering a particular family of graphs.

Let $G$ be any graph.  For each vertex $\alpha$, let $m_\alpha$ be some positive integer.  Consider a graph $\tilde G$ obtained by replacing each vertex $\alpha$ of $G$ by $m_\alpha$ copies of that vertex.  Label these copies by a pair $\alpha,a$, where $a\in 1,\ldots,m_\alpha$.  There is an edge in $\tilde G$ between vertices $\alpha,a$ and $\beta,b$ iff there is an edge in $G$ between $\alpha$ and $\beta$.

Note that for any $\alpha,a,b$, the operators $O_{\alpha,a} O_{\alpha,b}$ commute with all other operators and they square to $+1$.  So, we may assume that $O_{\alpha,a}=\pm O_{\alpha,b}$.  Without loss of generality, indeed we may assume that $O_{\alpha,a}=O_{\alpha,,b}$ (since we can pick coefficients $J_{\alpha,a}$ with arbitrary sign).
So, indeed, we may take simply $O_{\alpha,a}=O_\alpha$.

Then, the optimum for $\Psi(\tilde G)$ clearly has $J_{\alpha,,a}=J_{\alpha,b}$ for all $\alpha,a,b$, so we simply write $J_{\alpha,a}=\frac{1}{m_\alpha} K_\alpha$ for scalars $K_\alpha$.
Thus we have that
$\Psi(\tilde G)$ is the maximum, over all $K_\alpha$ such that
$$\sum_\alpha \frac{K_\alpha^2}{m_j}=1,$$
of
the square of the operator norm of $\sum_\alpha K_\alpha O_\alpha$.

Note that if we allow the $m_\alpha$ to be arbitrary positive real numbers, this defines a weighted problem on $G$, which we term the ``weighted $\Psi$ function of the graph".
The weighted $\Psi$ function is lowered bounded by the maximum, over independent sets of $G$, of the \emph{weight} of that independent set, where the weight is defined to be the sum of $m_\alpha$ in that set.  Indeed, given any independent set $S$ with weight $w$,
for $\alpha$ in that set
take $K_\alpha=m_\alpha/\sqrt{w}$ and then $\sum_\alpha \frac{K_\alpha^2}{m_\alpha}=1$, while the operator norm squared of $\sum_\alpha K_\alpha O_\alpha$ is $w$.

Now, suppose one succeeds in finding a large separation between the weighted $\Psi$ function and weighted independence number for some weighted graph $G$.  Does this imply that for some unweighted graph there is a large separation between $\Psi$ and the independence number?  It does.  If the weights are integers, we apply the construction above to obtain the $\tilde G$.  If the weights are rationals, we multiply them by a constant so that they become integer.  If the weights are real numbers, we approximate by rationals to any desired accuracy.

Note also that in general the optimum choice of coefficients $K_\alpha$ must be, by calculus, such that
$$\langle O_\alpha \rangle =\lambda \frac{K_\alpha}{m_\alpha},$$
where $\lambda$ is a Lagrange multiplier and where $\langle \ldots \rangle$ denotes the expectation value of the operator in any eigenstate of $H=\sum_\alpha K_\alpha O_\alpha$ of maximal eigenvalue.  Indeed, this equation must hold for all eigenstates of $H$ of maximal eigenvalue.

This potentially gives another way to construct improved examples with larger separation.  Take any example, such as our example above, which gives some separation even with all coefficients $J_\alpha$ equal to each other.  Then, either adjust the coefficients $J_\alpha$ to maximize the largest eigenvalue of $\sum_\alpha J_\alpha O_\alpha$, or else keep the coefficients fixed and equal to each other, and adjust the $m_\alpha$ so that this equation holds.

\subsection{Rounding}
We have conjectured that $\Psi(G)$ may be arbitrarily larger than $\alpha(G)$.  Conversely, one may ask whether $\vartheta(G)$ may be arbitrarily larger than $\Psi(G)$.  We conjecture that this is possible but do not have a candidate class of examples.

However, let us remark on the trivial fact that $\vartheta(G)$ may be strictly larger than $\Psi(G)$.  Indeed, this happens (as noted in Ref.~\onlinecite{hastings2022optimizing}) for the case that $G=C_5$ where $\vartheta(G)=\sqrt{5}$ but $\Psi(G)=\alpha(G)=2$.
This is interesting for the following reason.  Suppose we have indeterminants $\chi_\alpha$ obeying the relations that if $\alpha,\beta$ are neighbors then $\chi_\alpha \chi_\beta=-\chi_\beta \chi_\alpha$ while if $\alpha,\beta$ are not neighbors in $G$ then $\chi_\alpha \chi_\beta=\chi_\beta \chi_\alpha$.  However, suppose we do \emph{not} impose the relations $\chi_\alpha^2=1$, but only impose that $\sum_j\alpha\expec[\chi_\alpha^2]=n$, so we only impose the constraint in average in expectation value.
Consider some $H=\sum_\alpha J_\alpha \chi_\alpha$, with $\sum_j J_\alpha^2=1$.
Then, the degree-$2$ sum-of-squares bounds $\Vert H^2 \Vert \leq \vartheta(G)$, but no higher order provides any further bound (since we no do not have the relation $\chi_\alpha^2=1$).  However, we can obtain a representation of operators $\chi_\alpha$ obeying these relations by letting
$$\chi \equiv b_\alpha \xi_\alpha,$$
where $b_\alpha$ are scalar random variables subject to $\sum_\alpha\expec[b_\alpha^2]=n$ and
$\xi_\alpha$ obey the relations that $\alpha,\beta$ are neighbors then $\xi_\alpha \xi\beta=-\xi_\beta \xi_\alpha$ while if $\alpha,\beta$ are not neighbors in $G$ then $\xi_\alpha \xi_\beta=\xi_\beta \xi_\alpha$, and also they obey the relations that $\xi_\alpha^2=1$.
However, for any given $b_\alpha$, we may bound the operator norm squared of $\sum_\alpha J_\alpha \chi_\alpha=\sum_\alpha J_\alpha b_\alpha \xi_\alpha$ using the bound $\Psi(G)=2$.
Indeed, we find that
$$\Vert \sum_j J_\alpha \chi_\alpha\Vert^2 \leq 2 \sum_j b_\alpha^2,$$
which is equal to $2$ in expectation value and so is strictly smaller than the bound from sum-of-squares at degree $2$ or higher.

This is in contrast to the case of the degree-$2$ sum of squares for commuting variables.  If we have variables $\phi_i$ which commute with each other, and we constrain $\expec[\phi_i^2]=1$, but do not constrain $\phi_i^2=1$, then the degree-$2$ classical sum-of-squares will exactly compute the optimum of any quadratic function of these variables $\phi_i$.  This can then be used to round classical solutions by, e.g., replacing each random variable with its sign as in the Goemans-Williamson algorithm.

\section{SYK Model For $q\neq 4$}
\label{SYKqn4}
There is a proof\cite{feng2019spectrum} that, with high probability, for even $q\geq 2$ the operator norm of an instance of the SYK model is bounded by $O(\sqrt{n})$.
However, this proof does not give one a method of verifying\cite{hastings2022optimizing} for any \emph{specific instance} of the SYK model that the norm is $O(\sqrt{n})$.
For $q=4$, it was shown\cite{hastings2022optimizing} that the sum-of-squares could be used to give such a verification method; with high probability, a ``fragment" of the degree-$6$ sum-of-squares can bound the expectation value (in an arbitrary state) of an instance of the SYK Hamiltonian by $O(\sqrt{n})$.  Unfortunately, a straightforward generalization of that proof does not appear to give a tight (up to constant factors) bound for the norm for $q\neq 0,1,2,4$, where of course the $q=0,1,2$ cases are trivial.  The reader is invited to see this for themselves.

Let us show in \cref{oddqSYK} that, with high probability, for odd $q$ the operator norm of the SYK Hamiltonian is bounded by $O(1)$.  This proof will be done in an analogous manner to the proof of Ref.~\cite{feng2019spectrum}.  This proof may already exist in the literature, but I do not know a reference.

As explained in Ref.~\cite{hastings2022optimizing}, such a proof does not give a bound for any specific instance.
This raises the question: is there a proof method (ideally sum-of-squares, but perhaps some other proof) that, given an SYK instance, will, with high probability, generate an efficiently verifiable proof that the largest operator norm is $O(\sqrt{n})$ for even $q$ and $O(1)$ for odd $q$?  We have this proof method for $q=0,1,2,4$ but not for other $q$.

We remark that \emph{if} there is such a sum-of-squares proof for some given odd $q'$, then we have such a sum-of-squares proof for $q=q'+1$ which is even.  Consider the SYK model for some even $q$.  We can write $H=\sum_i \gamma_i \tau_i$, where each $\tau_i$ is a random instance of the odd $q'=q-1$ SYK model, up to scaling by a factor proportional to $1/\sqrt{n}$.
Then, we have
$\expec[H^2] \leq \expec[\sum_i \gamma_i^2] \expec[\sum_i \tau_i^2]$ by Cauchy-Schwarz, which sum-of-squares ``knows".  Using the assumed proof for $\expec[\tau_i^2]$, we have $\expec[\sum_i \tau_i^2]=O(1)$,
while $\expec[\sum_i \gamma_i^2]=n$ giving the desired bound on $\expec[H^2]$ which implies a bound on $\expec[H]$.

\begin{lemma}
\label{oddqSYK}
With high probability, the operator norm of a random instance of the SYK model for odd $q$ is $O(1)$.
\begin{proof}
Define $$\tau_i\equiv \sum_{i_2,\ldots,i_q} J_{i,i_2\ldots,i_q} \gamma_{i_2} \ldots \gamma_{i_q}.$$
For $i,j,k$ all distinct, define $$\tau_{i,,k}\equiv \sum_{i_4,\ldots,i_q} J_{i,,j,k,i_2\ldots,i_q} \gamma_{i_4} \ldots \gamma_{i_q}. $$
Similarly define $\tau_{i,j,k,l,m}$ and so on, for $\tau$ with any odd number of subscripts, less than or equal to $q-1$ subscripts.
Then, $H=(q!)^{-1} i^{(q-1)/2} \sum_i \gamma_i \tau_i$.

Now consider $H^2=(1/2) \{H,H\}$ for odd $q$.  We can expand $H$ as a sum of terms, each being monomials in Majorana operators, and similarly we can expand the anti-commutator $\{H,H\}$ as a sum of anti-commutators of monomials.   Since $q$ is odd, the monomials are of odd degree, and so two monomials anti-commute unless they agree in an odd number of indices.  So, we can express the anti-commutator as a sum of terms where the monomials agree in $1$ index, plus a sum of terms where they agree in $3$ indices, and so on.
Expanding in this way, we find that
$H^2=c_1 \sum_i \tau_i^2 + c_3 \sum_{i,j,k} \tau_{i,j,k}^2 + c_5 \sum_{i,j,k,l,m} \tau_{i,j,k,l,m}^2+\ldots$,
where the coefficients $c_1$ are all $O(1)$ and can be determined by some combinatorics. 

For some guidance, note that for any given $i$, the quantity $\tau_i$ is equal to a constant times $1/\sqrt{n}$ times a random instance of the SYK model of degree-$(q-1)$.

Now consider $$\expec[{\rm tr}\Bigl((H^2)^m\Bigr)],$$ for arbitrary integer $m>0$, where the expectation value is over random instances of the SYK model.
The trace for any given $H$ is bounded by $\Vert H \Vert^{2m} d_{\rm Hilb}$, where $d_{\rm Hilb}$ is the dimension of the Hilbert space which is exponential in $n$.  
We will bound the expectation value of the trace for some $m$ proportional to $n$ by an exponential of $n$, which will then imply the bound on $\Vert H \Vert$.

To bound the expectation value of the trace, expand each occurrence of $H$ as a sum of monomials in Majorana operators, and so in turn expand $H^{2m}$ as a sum of monomials of Majorana operators, with coefficients that are products of the coefficients of $\vec J$.  The trace of any given product of Majorana operators is either $0$ or $\pm1$, while the expectation value of any given product of $\vec J$ is non-negative.  Hence, we can upper bound the trace by replacing the trace of any product of Majorana operators by $+1$, giving
 $$\expec[{\rm tr}\Bigl((H^2)^m\Bigr)] \leq d_{\rm Hilb} \expec[\Bigl(c_1 \sum_i v_i^2 + c_3 \sum_{i,j,k} v_{i,j,k}^2 + \ldots\Bigr)^m],$$
 where $v_1,v_3,\ldots$ are random variables defined so that $v_i$ is the sum of coefficients (of monomials of Majorana operators) in $\tau_i$ and $v_{i,j,k}$ is the sum of coefficients in $\tau_{i,j,k}$, and so on.
 
 For each of these quantities, $\sum_i v_i^2$ and $\sum_{i,j,k} v_{i,j,k}^2$ and so on, we have a tail bound as follows.
 Consider the case of $\sum_i v_i^2$.  Each $v_i$ is a Gaussian random variable, with mean-square proportional to $1/n$.
 If these were independent random variables, the sum $\sum_i v_i^2$ would be chi-squared distributed and standard tail bounds are known.  They are not independent random variables, however, as coefficients $J_{i,j,\ldots}$ appear in both $v_i$ and $v_j$.  Thus, while $\expec[v_i^2]$ is proportional to $1/n$, we have that
 $\expec[v_i v_j]$ is proportional to $1/n^2$.
 However, we can still derive a tail bound.
Consider $\expec[\exp(a n \sum_i v_i^2)]$.  For sufficiently small $a>0$, this expectation value is finite, as one may calculate the Gaussian integrals and they converge.
So, the probability that $\sum_i v_i^2$ is greater than $c$ is exponentially small in $(c-c_0)n$ for some $c_0$.
Similar tail bounds follows for $\sum_{i,j,k} v_{i,j,k}^2$ and so on.
So, 
 we obtain a similar tail bound that the probability that the sum $c_1 \sum_i v_i^2 + c_3 \sum_{i,j,k} v_{i,j,k}^2 + \ldots$ is bigger than some $c'$ is in turn exponentially small in $(c'-c'_0)n$ for some $c'_0$.
This tail bound in turn 
bounds $\expec[\Bigl(c_1 \sum_i v_i^2 + c_3 \sum_{i,j,k} v_{i,j,k}^2 + \ldots\Bigr)^m]$ 
for some $m$ proportional to $n$ by an exponential of $n$, which completes the proof.
\end{proof}
\end{lemma}

\section{Quantum Knapsack Problem}
\label{qknapsack}
Here we describe a problem that we call the quantum knapsack problem.  First, let us give our motivation.  Consider a system of $n$ qubits, and some Hamiltonian which is a scalar plus linear and quadratic terms in the Pauli operators.  Then a degree-$2$ sum-of-square lower bound for the Hamiltonian means that we write
$$H=\sum_\alpha O_\alpha^\dagger O_\alpha+\lambda,$$
where each $O_\alpha$ is a scalar plus a term linear in Pauli operators.
Let us set the additive constant $\lambda$ to zero for simplicity.

In Ref.~\cite{hastings2023field}, two properties of this degree-$2$ decomposition were noted.  First, if we allow $O_\alpha$ to be non-Hermitian, then the sum-of-squares method is more powerful than if we restrict to proofs where $O_\alpha$ is Hermitian; indeed, using Hermitian $O_\alpha$, one does not even correctly compute some low order perturbation theory properties.  Second, it was noted that if we do restrict to Hermitian $O_\alpha$, then a sum-of-square proofs gives a decomposition of the Hamiltonian for the auxiliary field quantum Monte Carlo method where \emph{the severity of the sign problem depends on the difference between the true ground state energy and the sum-of-squares lower bound}.
These two properties mean that, roughly, Hermitian $O_\alpha$ generally do not bounds on the ground state energy which are close to the true energy, but if they do happen to give a close lower bound, then one may be able to improve using Monte Carlo methods.

So, the question naturally arises, suppose we use non-Hermitian $O_\alpha$, and suppose the lower bound from the degree-$2$ sum-of-squares is close to the true ground state energy, can we improve it?  One possibility is perturbing the solution of the semi-definite program\cite{hastings2022perturbation}.

However, let us just consider this as an abstract problem.  For simplicity, let us suppose the sum-of-squares method gives only a single term in the decomposition.  That is, suppose we have
$$H=O^\dagger O,$$ for some (possibly non-Hermitian) $O$ which is a scalar plus a term linear in Pauli operators.
We ask for the difficulty of approximating the true ground state energy of $H$, to some accuracy, such as inverse polynomial or inverse exponential.
We will this problem the ``quantum knapsack problem", with various forms of the problem depending on the accuracy of approximation.   One may also consider the quantum knapsack problem given promises that the true 
 the true ground state energy is either polynomially or exponentially small (i.e., relevant to the case that the sum-of-squares lower bound is close to the true ground state energy).  

We call this the quantum knapsack problem for the following reason.  Suppose that $O$ happens to be Hermitian.  In that case, $O$ is a scalar, plus a sum of Pauli operators.  By a unitary rotation on each qubit, we may assume that $O$ is of the form $O=c+\sum_i a_i Z_i$ for real scalars $c,a_i$.  In this case, the minimum of $O^2$ is given by choosing each $Z_i$ to be either $\pm 1$ such that $c+\sum_i a_i Z_i$ is as close as possible to zero.  This problem is very similar to the standard subset sum problem, which itself is a special case of the knapsack problem.
If we allow $O$ to be non-Hermitian, then this give some interesting ``quantum" generalization of this problem.

If $O$ is non-Hermitian, then we may write $O=A+iB$ for Hermitian $A,B$.  In this case,
$O^\dagger O=A^2 + B^2 + i[A,B].$  In the special case that $A,B$ commute, then we can apply a unitary rotation so that 
 $O$ is of the form $O=c+\sum_i a_i Z_i$ for \emph{complex} scalars $c,a_i$.
 
 The most interesting case is that $A,B$ do not commute.  We now give an approximation algorithm that can be useful if the true ground state energy is super-exponentially small.
 It is well known that given any matrix (Hermitian or not), one can apply a unitary rotation to bring it to lower triangular form.  Write $O=\sum_i O_i+c$ where $c$ is a complex scalar and each $O_i$ is a (possibly non-Hermitian) sum of Pauli operators on qubit $i$.
 Then, we may apply independent unitary rotations to each qubit so that each $O_i$ becomes of the form $O_i=c_i Z_i +d_i A_i$, where
 $$A_i\equiv \begin{pmatrix} 0 & 0 \\ 1 & 0 \end{pmatrix},$$
 and where $c_i,d_i $ are complex scalars.
 
 Label the qubits $1,\ldots,n$ and
 let $O_k$ be the $2^k$-by-$2^k$ matrix corresponding to $c+\sum_{i=1}^k O_i$.  Then,
 $$O_n=\begin{pmatrix} O_{n-1}+c_n & 0 \\ d_n & O_n-c_n \end{pmatrix}.$$
 
 Let $H_k=O_k^\dagger O_k$.
 The eigenvalues of $H_k$ are the squares of the singular values of $O_k$.  Let $\lambda^\uparrow_{n-1}$ be the least singular value of $O_{n-1}+c_n$
 and let $\lambda^\downarrow_{n-1}$ be the least singular value of $O_{n-1}-c_n$.  It is immediate that the least singular value of
 $$M_n\equiv \begin{pmatrix} \lambda^\uparrow_{n-1} & 0 \\ d_n & O_{n-1}-c_n \end{pmatrix}$$ is less than or equal to the least singular value of $O_n$, as
 acting on any right singular vector of $O_n$ the result is the same in the second block and has possibly reduced $\ell_2$ norm in the first block.
 To compute the singular values of $M_n$ we consider
 $$M_n^\dagger M_n=\begin{pmatrix} (\lambda^\uparrow_{n-1})^2+|d_n|^2 & \overline d_n (O_{n-1}-c_n) \\ d_n (O_{n-1}-c_n)^\dagger & (O_{n-1}-c_n)^\dagger (O_{n-1}-c_n)\end{pmatrix}.$$
By working in an basis of singular vector of $O_{n-1}-c_n$, we see that we may lower bound the singular values by the square-root of the eigenvalues of the two-by-two matrix
$$\begin{pmatrix} (\lambda^\uparrow_{n-1})^2+|d_n|^2 & \overline d_n\lambda^\downarrow_{n-1} \\ d_n \lambda^\downarrow_{n-1} & (\lambda^\downarrow_{n-1})^2\end{pmatrix}.$$
This gives a lower bound (for $\lambda^\uparrow_{n-1}   \lambda^\downarrow{n-1} \ll |d_n|$) by
$$ \frac{\lambda^\uparrow_{n-1} \lambda^{\downarrow}_{n-1}}{|d_n|}-o(\frac{\lambda^\uparrow_{n-1} \lambda^{\downarrow}_{n-1}}{|d_n|}).$$
For $\lambda^\uparrow_{n-1}\ll  |d_n|$ and $\lambda^\downarrow_{n-1}\gtrsim |d_n|$, this is lower bounded by a constant times $\lambda^\uparrow_{n-1}$, and similarly with $\uparrow,\downarrow$ interchanged.
If $\lambda^\uparrow_{n-1}\gtrsim |d_n|$ and $\lambda^\downarrow_{n-1}\gtrsim |d_n|$, then this is lower bounded by a constant times $({\rm min}(\lambda^\uparrow_{n-1},\lambda^{\downarrow}_{n-1})$.

Suppose all $|d_k|$ are $O(1)$.  
Let us use $\vec \sigma$ to denote a basis vector in the $Z$ basis, i.e., an assignment of $\uparrow$ or $\downarrow$ for each of the $n$ qubits.
Let $\lambda_{\vec \sigma}$ denote the corresponding eigenvalue of $|c+\sum_i c_i Z_i|$; note the absolute value sign.
Then, applying this recursively (to lower bound $\lambda^{\uparrow}_{n-1}$ and $\lambda^{\downarrow}_{n-1}$ in terms of operators on $n-2$ qubits, and so on), we find that
the singular values of $O$ are lower bounded by the \emph{product} of $\lambda_{\vec \sigma}/O(1)$ over all $\vec \sigma$.
Let $\lambda_{\rm min}$ denote the minimum of $\lambda_{\vec \sigma}$ over all $\vec \sigma$.
Hence,
the lowest singular value of $O$ is lower bounded by $(\lambda_{\rm min}/O(1))^{2^n}$.  In particular, if the lowest singular value of $O$ is zero, then some $\lambda_{\vec \sigma}$ is zero.

We leave finding improvements to this approximation as a challenge.  \emph{What is the complexity of approximately solving this quantum knapsack problem, especially in the limit that the true ground state energy of $H$ is small?}

\bibliography{qsos-ref}
\end{document}